\documentclass[pra,aps,twocolumn,superscriptaddress,longbibliography]{revtex4-1}
\usepackage{graphicx,color,times,float,braket,url,amsmath,amssymb,comment}

\usepackage[colorlinks=true,citecolor=blue,linkcolor=blue,urlcolor=blue]{hyperref}

\begin{document}

\title{Localization with non-Hermitian off-diagonal disorder}

\author{Aitijhya Saha}
\affiliation{Harish-Chandra Research Institute, A CI of Homi Bhabha National Institute, Allahabad 211019, India}
\affiliation{Department of Theoretical Physics, Tata Institute of Fundamental Research, Mumbai 400005, India}
\author{Debraj Rakshit}
\affiliation{Harish-Chandra Research Institute, A CI of Homi Bhabha National Institute, Allahabad 211019, India}

\date{\today}

\begin{abstract}
In this work, we discuss a non-Hermitian system described via a one-dimensional single-particle tight-binding model, where the non-Hermiticity is governed by random nearest-neighbour tunnellings, such that the left-to-right and right-to-left hopping strengths are unequal. A physical situation of completely real eigenspectrum arises owing to the Hamiltonian's  tridiagonal matrix structure under a simple \emph {sign conservation} of the product of the conjugate nearest-neighbour tunnelling terms. The off-diagonal disorder leads the non-Hermitian system to a delocalization-localization {\color{black}crossover in finite systems}. The emergent nature of the {\color{black}crossover} is recognized through a finite-size spectral analysis. {\color{black}The system enters into a localized phase for infinitesimal disorder strength in the thermodynamic limit.} We perform a careful scaling analysis of localization length, inverse participation ratio (IPR), and energy splitting and report the corresponding scaling exponents. \textcolor{black}{Noticeably, in contrast to the diagonal disorder,  the density of states (DOS) has a singularity at $E=0$ in the presence of the off-diagonal disorder and the corresponding wavefunction remains delocalized for any given disorder strength.}
\end{abstract}
    
\maketitle

\section{Introduction}
The development of elementary quantum mechanics has been exclusively based on the fact that any physical observable should have real expectation value. This requirement is fulfilled when the Hamiltonian under consideration is Hermitian. On the other hand, non-Hermitian Hamiltonians do not have a complete and orthonormal eigenbasis, which leads to complex expectation values of physical observables. However, the existence of non-Hermitian Hamiltonians with real eigenvalues has been known for some time \cite{Bender98,Konotop16,Bergholtz21}. Non-Hermitian systems are essential for understanding a wide range of quantum phenomena. For example, non-Hermitian quantum dynamics addresses open quantum systems from a fresh perspective, and hence, appears quite frequently in the context of open dynamics \cite{Carmichael93,Xu16,Kawabata17,Lee19}. Here, one can mention the dynamics of continuously monitored quantum many-body systems that can be followed via non-Hermitian treatment, where the non-Hermitian formalism successfully explains individual quantum trajectories with no quantum jumps \cite{Minganti20,Fleckenstein22,Ashida20,Abbasi22,Agarwal05}. In general, non-Hermitian formulations of quantum dynamics provide effective descriptions in various contexts, that include topological systems \cite{Rudner102,Zeuner15,Lee16,Xu17}, quantum transport \cite{Giusteri15,Zhang19,Lima23}, nuclear phenomena \cite{Gamow28,Feshbach58,Feshbach62}, optics \cite{Makris08,Klaiman08,Ruter10}, quantum fields \cite{Fisher78}, among others. Importantly, there are various physical platforms via which non-Hermitian systems can be accessed in experiments \cite{Ozdemir19,Chen21}.

A prominent interest in physics beyond the Hermitian system began in the community with the systems not obeying the stringent condition of Hermiticity, but obeying certain symmetries, e.g., \textcolor{black}{parity-time} symmetry, known as $\mathcal{PT}$-symmetry \cite{Bender98,Bender02,Modak21}, or \textcolor{black}{rotation-time} symmetry, known as $\mathcal{RT}$-symmetry \cite{Zhang13,Lange10}. Such systems are shown to have two distinct phases -- The phase, where the entire eigen spectrum is real, the so-called $\mathcal{PT}$-unbroken phase in the context of \textcolor{black}{parity-time} symmetry, and another phase,  the broken phase, where the eigenspectrum can be partially or completely imaginary. The demarcation of these two phases occurs through a transition point, which is known as the exceptional point. Clearly, the underlying symmetries of the system are not respected in the broken phase. In fact, physical situations with completely real eigenvalues can arise with even lesser constraining conditions that are devoid of any above-mentioned symmetries. There, the situation can be as drastic as a non-symmetric  Hamiltonian comprising random matrix elements that admits an entirely real spectrum. In such a situation, the randomness leads to a physical situation of the well-known quantum phenomenon of localization, which is in spirit parallel to the famous Anderson localization.

Randomness naturally appears in physical systems. Given its ubiquitous nature and often non-intuitive effects it has on the system properties, disordered systems have been a subject of intense research for a long time \cite{Binder86,Belitz05,Das08,Alloul09}. There are plethora of disorder-induced nontrivial quantum phenomena, such as novel quantum phases \cite{Chowdhury86,Mezard87,Sachdev99,Yao14,Zuniga13}, order-from-disorder \cite{Aizenman89,Wehr06,Bera14,Bera16,Bera17}, high-$T_c$ superconductivity \cite{Auerbach94}, and localization \cite{Anderson58,Abrahams79,Evers08,Nandkishore15,Abanin19}. The fundamental phenomena of disorder-induced Anderson localization, which states that the presence of randomness may suppress the diffusion of waves \cite{Anderson58,Abrahams79}, take the central stage and are a well-established, still thriving research topic in modern condensed matter physics. Detailed analytical and numerical studies have been performed for analyzing different Anderson types of transitions, for developing their scaling theories, and for the classifications of their universality classes \cite{Hikami81,Altland97,Wang21,Luo21,Luo212,Luo22,Bu22}. The quantum phenomena of localization bear a wide impact given the experimental success of realizing it in synthetic materials \cite{Aspect09,Fallani08,Modugno10,Negro03,Roati08,Lahini09}, and due to further potential that it brings as a useful resource in quantum technology \cite{Mishra16,Sadhukhan15,Sadhukhan152,Sahoo23}.

While disordered Hermitian systems drew major attention in the community, as expected, some interesting ideas have also been pitched in for the disordered non-Hermitian systems as well \cite{Jiang19,Longhi19,Kawabata21,Weidemann22}. It began with the seminal work by Hatano and Nelson \cite{Hatano96,Hatano98}, who investigated a single-particle tight-binding model with asymmetric hopping and random onsite potential. The system is characterized via a real-complex transition -- The transition is marked by the destruction of the localization due to strong non-Hermiticity. Subsequently, this work was followed by others with interesting results \cite{Hatano96,Hatano98,Zeng20,Liu20,Liu202,Li20,Liu21,Lee21,Molignini,Padhan}, such as the presence of mobility edge in lower-dimensions. Primarily, these work concentrate on random or quasiperiodic onsite potentials, which are responsible for the localization, and a Hatano-Nelson type asymmetric hopping that imposes a symmetry on the system of the form, $\mathcal{H}(h)^T = \mathcal{H}(-h)$, where $h$ is associated with parameterization of the tunnelling parameter of the Hamiltonian, $\mathcal{H}$.

In this work, we consider a one-dimensional single-particle tight-binding non-Hermitian model with random nearest-neighbour tunnellings, such that  $t_{i,i+1} \ne t_{i+1,i}$, which makes the system non-Hermitian. The corresponding Hamiltonian describing the system has a tridiagonal structure. When the sign of the product of the conjugate pair of nearest-neighbour hopping terms is kept conserved, the matrix is characterized by entirely real eigenvalues. The reality of the eigenvalues in such systems is revealed through an existing mathematical theorem. Then, despite the non-Hermiticity in the system, a physical quantum phenomenon of localization emerges due to the randomness in the tunnelling parameter. 

\textcolor{black}{While the majority of the works are conducted in the context of localization, induced via random onsite potential, i.e., via diagonal disorder, there have been attempts with off-diagonal disorder-induced localization as well.  The effects of off-diagonal and diagonal disorder are assumed to be somewhat similar.  In one dimension, the striking difference between the diagonal and off-diagonal disorder is that for diagonal disorder, all the wavefunctions are localized, whereas for off-diagonal the wavefunction corresponding to the exact middle of the energy band remains delocalized for any amount of disorder \cite{Delyon83, Delyon87, Economou77,Ziman82,Inui94,Theodorou75, Soukoulis82}.}  Along with lower-dimension, these works address higher-dimensional systems as well, primarily, via approximate theories, and yield useful insights on the off-diagonal disorder-induced localization mechanism. However, they do not provide any commentary on setting up a link between the non-Hermiticity of the Hamiltonian and the reality of the eigenvalues. Thus, the results may appear accidental to an attentive reader. Ref.~\cite{Biswas2000}, which reports localization in the presence of random nearest neighbour hopping as well as random on-site potential, does not mention any constraint on the non-Hermitian Hamiltonian to have all-over real spectra, and the results are reported in the regime of complex eigenvalues also. Naturally, there is a lot of room for conducting rigorous studies in such systems, e.g., precise estimation of the scaling exponents dictating the nature of the localization-delocalization \textcolor{black}{crossover} for an off-diagonal disordered system. Given the current impetus with non-Hermitian systems, establishing the theories with off-diagonal randomness becomes relevant. In this work, apart from spectral statistics and DOS analysis, we perform careful finite-size scaling analysis in order to characterize the \textcolor{black}{crossover in the finite systems.} Suitable quantities, such as localization length, IPR, and the energy gap, are studied, and associated scaling exponents controlling their behavior \textcolor{black}{near the crossover} are reported.

The rest of the paper is arranged in the following manner. This introduction is followed by the model under study in Sec.~II. Thereby, we discuss the nature of spectral statistics both for the band edge and the mid-band in Sec.~III. In Sec.~IV we provide specific attention to the mid-band states via the study of DOS and localization length. Characteristic observables such as localization length, IPR, and energy gap between the ground and the first excited states, which are used for characterizing the finite-size scaling properties, are introduced in Sec.~V, where we report the respective scaling exponents. Finally, we provide a summary of the main findings, following concluding remarks in Sec.~VI.

\begin{figure*}[t]
  \includegraphics[width=\linewidth]{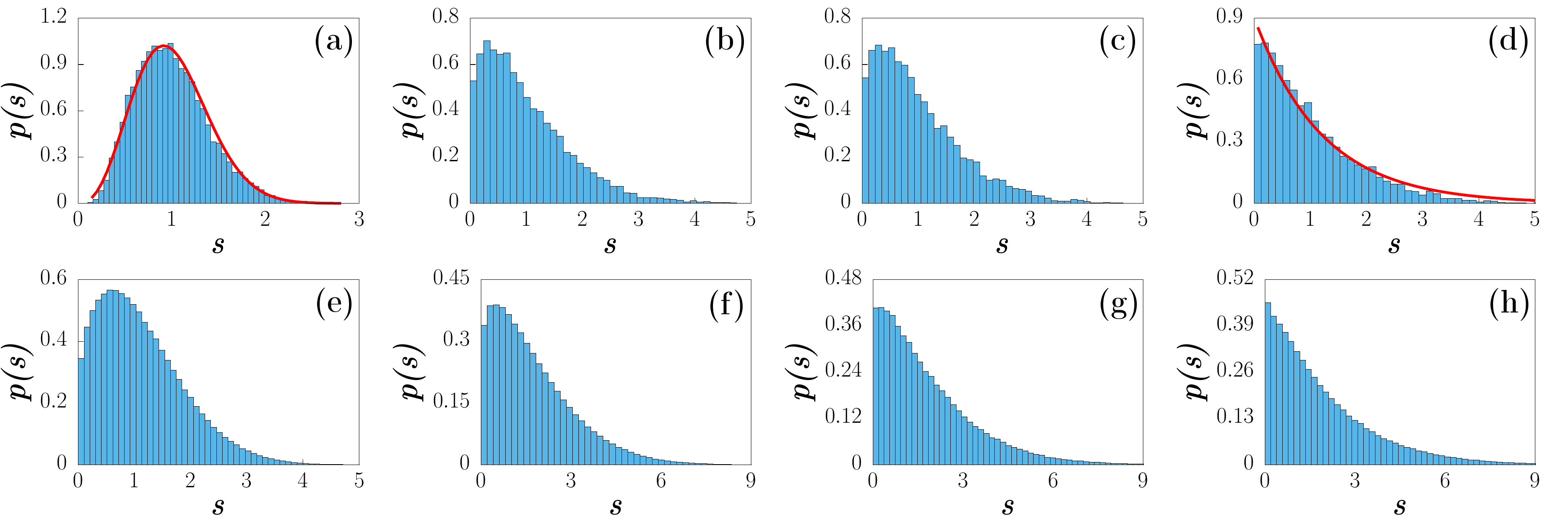}
  \caption{{\bf (a-d) Spectral Analysis at Band Edge:} (a) represents a case of distribution for the level spacing $s$ between the ground and first excited state for $L=50$ and $W=0.1$. Fitting a function of the form given in Eq.~\ref{Wigner-Dyson}, it turns out $\alpha=1.5$ and $\beta=2.5$, implying a Wigner-Dyson type distribution. Keeping the same disorder $W=0.1$ for $L=200$, the distribution shifts towards the origin as shown in (b). The same nature occurs in (c) for $L=50$ with $W=1.0$, which indicates that the spectral statistics move away from the Wigner-Dyson type distribution as we increase either the system size or the disorder. In the case of (d), where $L=200$ and $W=1.0$, it has become a Poisson type distribution having $\lambda=0.83$ as found by fitting a function of the form provided in Eq.~\ref{Poisson}. \textcolor{black}{{\bf (e-h) Spectral Analysis at Mid-Band:} In the mid-band, the wavefunctions being least localized, for smaller system sizes, the spectral statistics is indeed Wigner-Dyson type even for $W=1.0$, and consequently, for $L=200$ with $W=1.0$ in (e), the spectral statistics is still Wigner-Dyson type. For the same disorder, with increasing system size $L=500$, $1000$, $2000$ in (f), (g), and (h), respectively, the nature of the spectral statistics slowly moves towards a  Poisson-type distribution. This trend indicates that for all the energy levels, in the thermodynamic limit, the spectral statistics would collapse from Wigner-Dyson to Poisson-type distribution for an infinitesimal disorder.}}
  \label{Spectral_Analysis}
\end{figure*}

\section{Model}
We consider a Hamiltonian of the form,
\begin{equation}
    H=-\sum_{i=1}^{L-1}(t_{i+1,i}\,c^{\dagger}_{i}c_{i+1}+\,t_{i,i+1}c^{\dagger}_{i+1}c_{i}),
    \label{model}
\end{equation}
where $t_{i+1,i}$ and $t_{i,i+1}$ denote the random hopping strengths from right-to-left and from left-to-right, respectively. Here $c_{i}$ ($c^{\dagger}_{i}$) is the annihilation (creation) operator at the $i^{\mathrm{th}}$ lattice site.

In the case of Anderson localization, the diagonal terms representing the disordered potential have a form $W y$, where $W$ controls the strength of disorder and $y$ is drawn randomly from a uniform distribution, $y\sim U(-1,1)$. This random potential term, when weighted against the off-diagonal tunnelling terms, brings a comparative sense between the diagonal and off-diagonal terms, which in turn decides the relative strength of the disorder that the system is subjected to. In this work, we do not consider a potential term, and in a way, this relative sense is lost if $t$ is just proportional to $W$, and as a result, the emerging physics remains largely structureless.  This issue can be circumvented, if, instead, we allow $t$ to be a generic function of $W x$, i.e., $t = F(W x)$, where $x$ is just a random variable following a uniform distribution and $F$ is an arbitrary function having the characteristics that for $W\!\to\!0$, $t$ is fairly constant, whereas an increase in $W$ results in enhancement of the effects of randomness in $t$, but this increase in randomness is not proportional to a certain power of $W$.

In our case, we randomize the hopping strength such that the nearest-neighbor tunnelling probability is uncorrelated in space. We allow the maximum hopping strength to be unity. The particular form of the hopping term considered in this work is given by,
\begin{equation}
    t=(1+W x)^{-1},
    \label{hop}
\end{equation}
where $x$ is drawn randomly from a uniform distribution over the range zero to unity, i.e., $x\sim U(0,1)$, and $W$ parameterizes the amount of disorder in the system. For the small value of $W$, tunnelling is nearly the same at every lattice site and is only weakly affected by randomness. As the value of $W$ increases, the effects of randomness increase, and the number of inter-site hopping with comparatively weaker tunnelling strength increases. Here, we mostly concentrate on the small disorders as large disorders would cease hopping in most of the lattice sites, and the system behaves as an insulator. To model this kind of disorder, one can come up with a variety of functional forms; Eq.~\ref{hop} is one of those. To get rid of the statistical fluctuations in the results corresponding to various quantities under study, we perform standard quenched averaging over many realizations, i.e., we compute the value of the physical observable with a fixed configuration of the disorder, and finally, a configuration averaging of the quantity over many such realizations is performed. In this work, we report the results for an ensemble average of over 10000 samples. Here, we have used the open boundary condition, which makes the Hamiltonian under consideration tridiagonal.

Evidently, the Hamiltonian under consideration is non-Hermitian in nature. However, due to its tridiagonal form, it is found that the spectrum is always real, irrespective of the strength of disorder $W$. Let us consider a non-Hermitian tridiagonal matrix $\mathcal{A}_{NH}$ of order $N \times N$ with diagonal matrix elements as $a_i$, where $i \in [1, N]$, along with upper sub-diagonal and lower sub-diagonal matrix elements as $b_i$ and $c_i$, respectively, where $i \in [1, N\!-\!1]$ and in which the sub-diagonal elements assume non-zero real values, then
the following theorem is satisfied  \cite{Bohigas13, Marinello16}:
\begin{itemize}
    \item[] \textbf{Theorem:} If the products $b_i c_i\!>\!0$, $\forall i \in [1, N\!-\!1]$,  then all eigenvalues of $\mathcal{A}_{NH}$ are real. The matrix $\mathcal{A}_{NH}$ is pseudo-Hermitian.
\end{itemize}
In such cases, the non-Hermitian tridiagonal matrix $\mathcal{A}_{NH}$ can be mapped to a Hermitian tridiagonal matrix $\mathcal{A}_{H}$, whose diagonal elements are $a_i$'s and the sub-diagonal elements are $\sqrt{b_i c_i}$'s, via a certain suitable similarity transformation. 

\textcolor{black}{In our case, all the $a_i$'s are zero and consequently under the operation $\psi^n_j \to (-1)^j\psi^n_j$, i.e., if we change the sign of the coefficient of the wavefunction at alternating sites, the Hamiltonian changes its sign which manifests the fact that, corresponding to an energy eigenvalue $E_n$ there should exist another eigenvalue $-E_n$.} 

\section{Spectral Statistics}
\textcolor{black}{Particle transport and thermalization are seized in the localized phase, and thus a localized quantum many-body (QMB) system can evade ergodicity, which is, otherwise, the standard behavioral aspect that the system ought to exhibit in accordance with the well-known eigenstate thermalization hypothesis (ETH) \cite{Rigol08, Rigol16, Deutsch18, Huse15, Zelevinsky16}.} The statistical features of the spectrum of a QMB system in the localized and in the ergodic phases are expected to be markedly different. Thus, it is known to provide crucial insights into quantum ergodicity, as well as localization. This spectral statistics-based analysis is rooted in two basic conjectures, namely the Bohigas conjecture \cite{Bohigas84} and the Berry-Tabor
conjecture  \cite{Berry77}. Bohigas conjecture indicates that the systems with chaotic classical limits are equivalent to an ensemble of random matrices, and consequently, can be described by random matrix theory (RMT) \cite{Dyson621,Dyson622,Dyson623,Dyson72}, whereas Berry-Tabor conjecture predicts Poisson statistics for systems having an integrable classical limit. In particular, if a quantum chaotic Hamiltonian possesses time-reversal symmetry, it is expected to belong to the Gaussian Orthogonal Ensemble (GOE). Spectral statistics thus provide an effective means to classify chaotic and integrable systems. These conjectures can further be utilized for predicting the possible ergodic or localized nature of a QMB system. Quantum ergodicity is associated with random matrix theory, where the spectral distribution obeys Wigner-Dyson level statistics. Conversely, Poissonian level statistics are identified with localization. The relationship between the Poissonian level statistics and Anderson localization has been established concretely \cite{Shklovskii93,Kravtsov94,Aronov95,Kravtsov95,Basko06,Evers08}. 

In this work, we investigate the spacing between consecutive energy levels, which is a standard practice in the analysis related to spectral statistics. First, we consider the energy splitting distribution between the ground state energy, $E_0$, and the $1$st excited state energy, $E_1$. The parameter under consideration is $s$, which is defined as
\begin{equation}
    s = \frac{E_1-E_0}{\langle E_1-E_0\rangle},
\end{equation}
where the average in the denominator is basically the ensemble average. Following the two fundamental conjectures discussed above, an ergodic system should exhibit a Wigner-Dyson type statistics, i.e.,
\begin{equation}
    p(s) \sim s^{\beta} \exp(-\alpha s^2),
    \label{Wigner-Dyson}
\end{equation}
where the constant $\beta$ determines the types of the ensemble, e.g., Gaussian Orthogonal Ensemble (GOE) ($\beta=1$), Gaussian Unitary Ensemble (GUE) ($\beta=2$), or Gaussian Symplectic Ensemble (GSE) ($\beta=4$), while
onset of localization would imply that the probability of a normalized spacing between $s$ and $s+ds$ would obey a Poisson-type distribution, i.e.,
\begin{equation}
    p(s) \sim \exp(-\lambda s).
    \label{Poisson}
\end{equation}

In order to obtain the energy levels of the single-particle case described via the Hamiltonian in Eq.~\ref{model}, we perform exact diagonalization. As the dimension of the Hilbert space for a single particle grows only linearly with the lattice size, large system sizes can be probed, in principle. However, due to the requirement of ensemble averaging over a large number of samples, we carry out the computation upto $L=2000$ for spectral statistics, while in the case of scaling analysis of localization length and other quantities, we can consider upto $L=1000$. 

In the presence of infinitesimal disorder in a finite-size system, it is well known that the spectral statistics follows almost a Gaussian-type distribution with the mean at unity and the standard deviation increases with an increase in system size \cite{Bohigas13}. Fig.~\ref{Spectral_Analysis}(a) to \ref{Spectral_Analysis}(d) clearly demonstrate an evolving trend in the distribution structure as the system size and the disorder strength are enhanced.  The system is identified with a strong level repulsion at small $s$ for weak disorder. It is characterized by the Wigner-Dyson type distribution in Fig.~\ref{Spectral_Analysis}(a) for a particular case with $W=0.1$ and $L=50$. It is observed that an increase in system size with the same disorder results in moving the peak of the distribution towards small $s$ values, which is shown in Fig.~\ref{Spectral_Analysis}(b) in the case where $W=0.1$ and $L=200$. It indicates the fact that the level repulsion decreases for large systems at even smaller disorders, which is quite obvious because an increase in system size brings more scope of randomness, and it is further justified in Fig.~\ref{Spectral_Analysis}(c) where $L=50$ but the randomness in hopping is being increased by taking $W=1.0$. Finally, for $W=1$ and $L=200$ in Fig.~\ref{Spectral_Analysis}(d), the distribution has become almost Poisson type, suggesting a localized nature of the wavefunction. \textcolor{black}{As the spectrum of this system is symmetric about the $E=0$ energy eigenvalue, both the low and high lying band edges show a similar kind of spectral statistics. Hence, we next focus on the mid-band spectral statistics.} 

\textcolor{black}{By doing a similar analysis, we find that for smaller system sizes, even for an appreciable amount of disorder, the mid-band spectral statistics, i.e., the spectral statistics of the energy gap between the two energy levels in the center of the band, remain Gaussian type. The Wigner-Dyson type distribution is observed for even larger systems. In Fig.~\ref{Spectral_Analysis}(e) to \ref{Spectral_Analysis}(h) the spectral statistics are shown for the disorder $W=1.0$ for the system sizes $L=200$, $500$, $1000$, and $2000$. It is easily noticeable that while the energy gap between the ground and 1st excited state shows a Poisson-type distribution for $L=200$ with $W=1.0$, the mid-band energy gap, i.e., the energy gap between the two central eigenenergies, still shows Wigner-Dyson type distribution. As before, with an increase in system size, the statistics move towards a Poisson-type distribution. So, in terms of spectral statistics, it can be inferred that for an appreciable amount of disorder and system size, all the states are localized, which is, however, not the case as we discuss below.}

\section{Mid-Band Anomaly}

\begin{figure}[t]
    \centering
    \includegraphics[width=\linewidth]{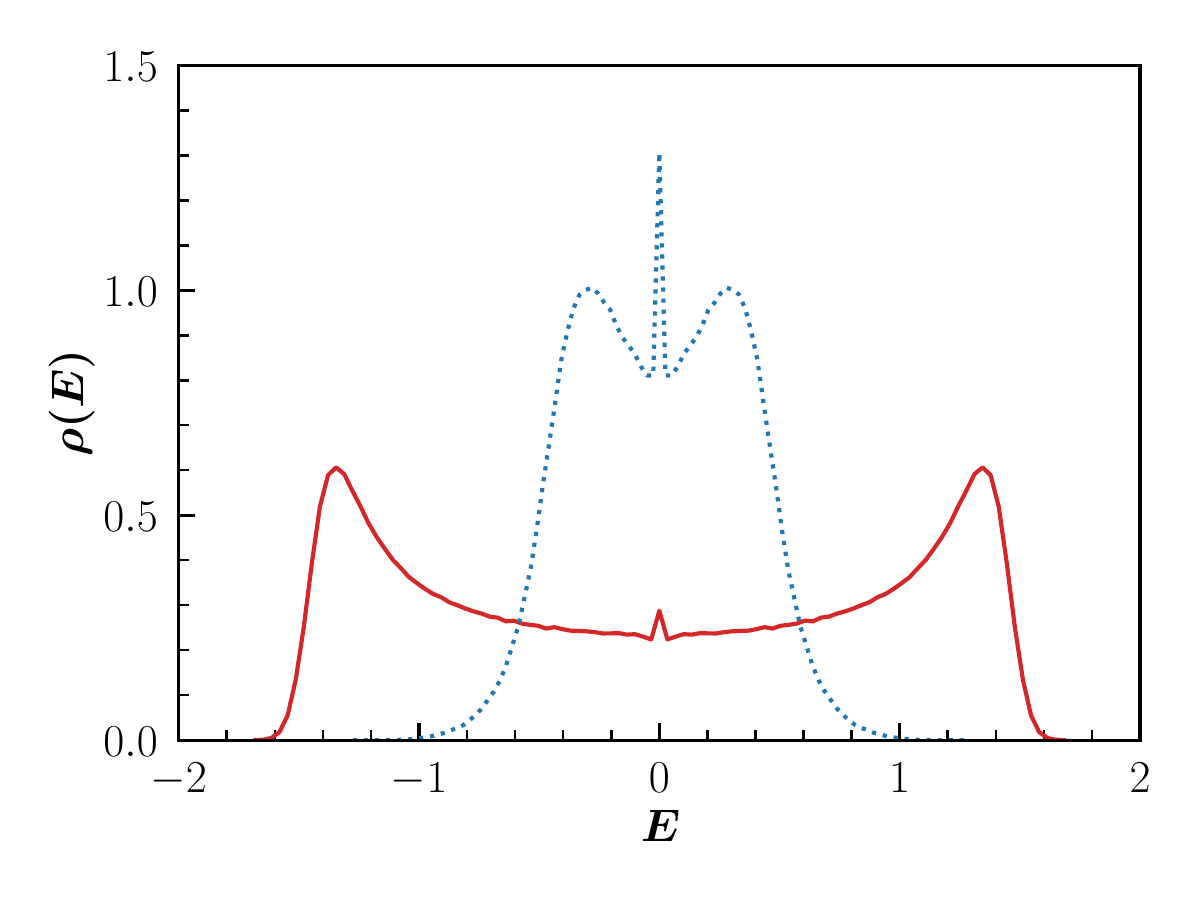}
    \caption{{\bf Density of States:} The red solid line and blue dotted line correspond to the density of states for disorder strength $W=1$ and $W=10$, respectively. An increase in the disorder results in a larger peak at $E=0$ energy. The density of states is normalized in such a way that $\int \rho(E)~dE=1$. The presence of singularity in the mid-band is related to the delocalized nature of the wavefunction.}
    \label{Density of States}
\end{figure}

\begin{figure}[t]
    \centering
\includegraphics[width=\linewidth]{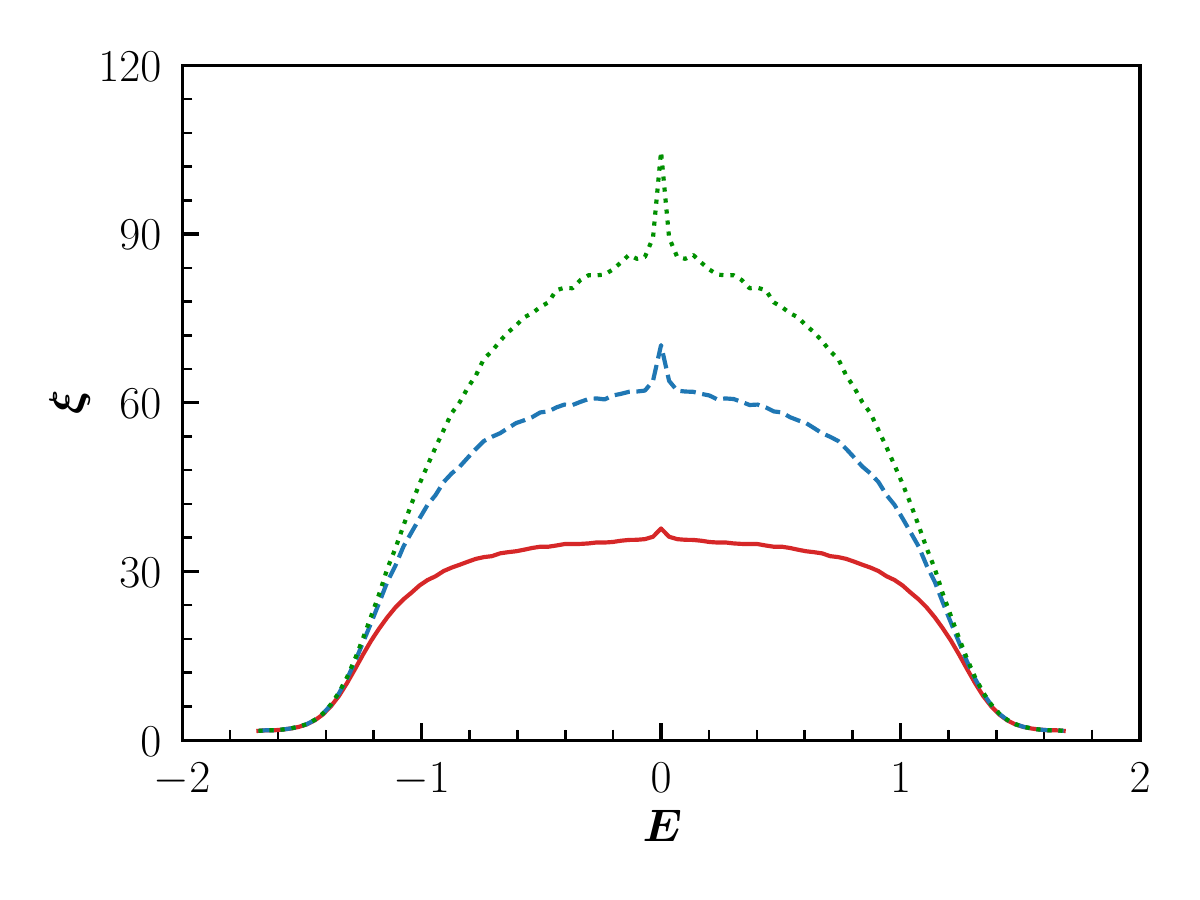}
    \caption{{\bf Localization Length vs Energy:} The red solid line, blue dashed line, and green dotted line correspond to system size $L=200$, $500$, and $1000$, respectively. For all of these, the disorder strength is kept fixed at $W=1.0$. In all three cases, there is a rise in the value of localization length at the middle of the spectrum, and this behaviour becomes stronger with an increase in system size. At the thermodynamic limit, only $E=0$ has a delocalized eigenstate.}
    \label{Zeta vs Energy}
\end{figure}

\textcolor{black}{In this context, we also investigate how the DOS varies with energy. From Fig.~\ref{Density of States} we find a prominent peak at the $E=0$ energy. This peak becomes sharper with an increase in disorder strength. We know that in the disorder-free case, in the thermodynamic limit, for nearest neighbour hopping in the tight-binding model, there exists a singularity in the band edge. The inclusion of disorder removes this singularity in the band edge. In the case of diagonal disorder, there appears to be no such singularity in the DOS, in contrast to off-diagonal disorder. For off-diagonal disorder, this singularity is related to the delocalized nature of the wavefunction at the mid-band.} To quantify the localized or delocalized nature of the wavefunction, we use the definition of localization length,
\begin{equation}
    \xi=\sqrt{\sum_{j=1}^{L}(j-j_0)^2p_j},
\end{equation}
where $j$ denotes the lattice site, $p_j$ is the single-particle probability density at site $j$ and $j_0$ is the localization center having the expression \textcolor{black}{$j_0=\sum_{j=1}^{L}jp_j$}. Expanding the $n^{\mathrm{th}}$ normalised eigenstate of the system, $|\psi^n\rangle$, in terms of the single-particle computational basis, $|j\rangle$, such that $|\psi_n\rangle = \sum_j c^{(j)}_n |j \rangle$, $p_j$ is given by $p_j=|\langle j|\psi^n\rangle|^2=|c^{(j)}_n|^2$. \textcolor{black}{To be specific, in this entire work, we have used the right eigenvector as the eigenstate. However, the same results are obtained if one uses the left eigenvector instead. From Fig.~\ref{Zeta vs Energy} we find that there is indeed a peak in the center of the band for the localization length, and this peak becomes even sharper with system size for a fixed amount of disorder. Following the trend, one may infer that there will remain a delocalized wavefunction at $E=0$ in the thermodynamic limit.}

\textcolor{black}{ There should be an associated finite-size scaling analysis to characterize the crossover. One can, in principle, perform this analysis for all the available states. However, traditionally, for Anderson localization or Aubry-Andr\'{e} (AA) localization, the research community is mainly interested in the scaling analysis for the ground state, which is quite justified in the case of quantum phase transition (QPT). Hence, we also concentrate on it and pursue a detailed study in the following section.}

\section{Scaling Analysis}
In the periodic lattice systems, the conventional second-order quantum phase transition is described via the fixed points of the renormalization group, according to which the correlation length, $\xi$, diverges at criticality as $\xi \sim |h-h_c|^{-\nu}$, where $h$ is a control parameter and  $h_c$ is the critical point. The critical exponent, $\nu$, dictates the rate of divergence. The divergence of the characteristic length scale of the system essentially makes the system scale-free, which, in turn, gives rise to universality at the criticality.

The translational symmetry of a system is lost in the presence of disorder or even in the absence of it, for example, in the presence of a quasi-periodic potential.  \textcolor{black}{ In general, the localization-delocalization transition does not belong to the Landau paradigm of second-order phase transition as it is not characterized by a spontaneous symmetry breaking at the transition point.}  However, a parallel scaling analysis can be performed in the case of localization transition as well, where the characteristic length scale associated with the system, $\xi$, can be identified to be the localization length with disorder strength, $W$, as the control parameter driving the transition. Considering $W_c$ as the critical point, the near-critical scaling of the localization length is given by $\xi \sim |W-W_c|^{-\nu}$.

\begin{figure}[t]
  \centering
\includegraphics[width=\linewidth]{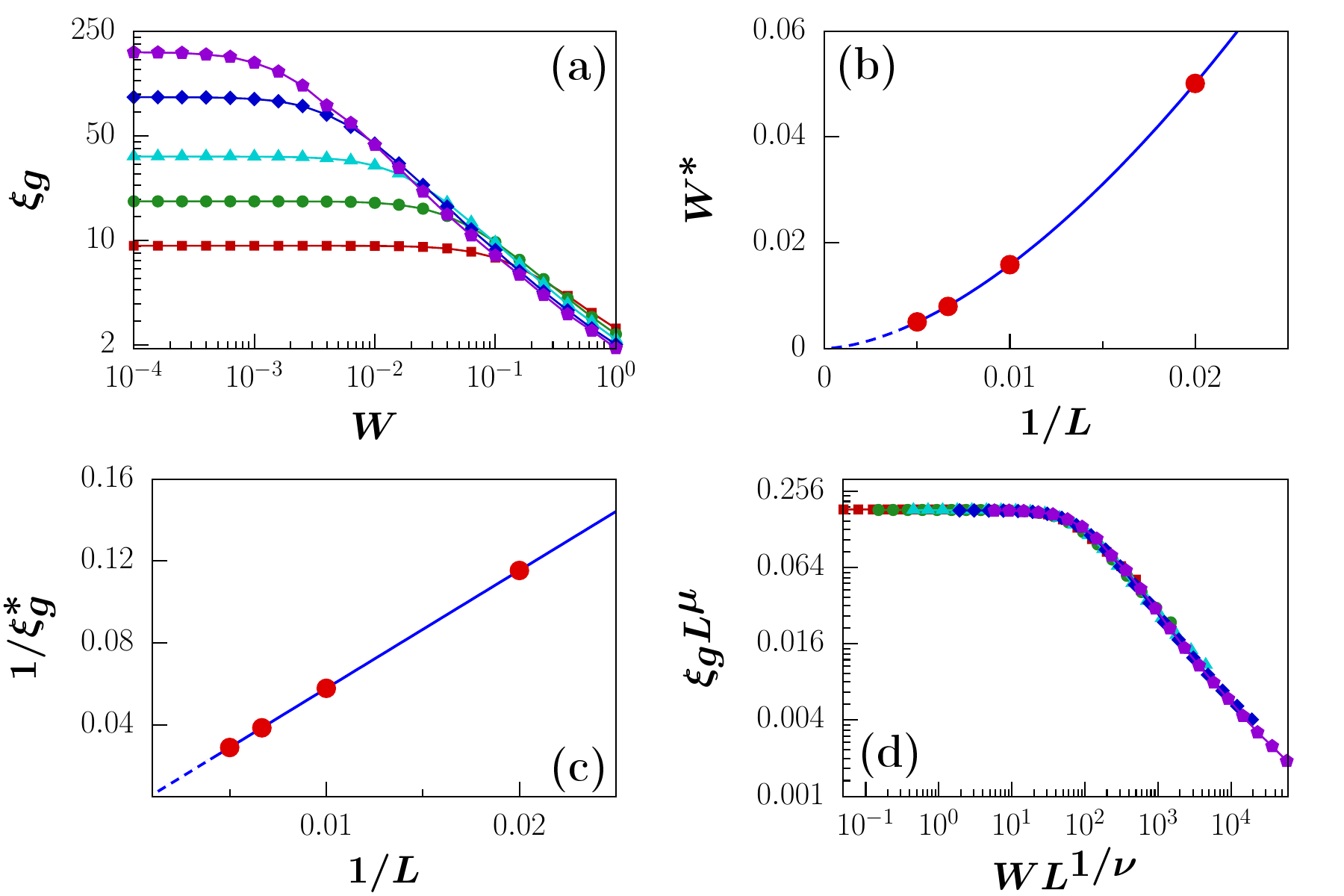}
    \caption{{\bf Localization Length:} (a) shows the variation in localization length of the ground state, $\xi_g$, with increasing disorder $W$ for various system sizes $L$, where square, circle, triangle, diamond, and pentagon represent the cases for \textcolor{black}{$L$ = 50, 100, 200, 500, and 1000} respectively. They are characterized by an initial flat region, where the corresponding wave function has an extended nature. Beyond a certain disorder $W^{*}$, the localization length has only a very weak dependence on the system size and decreases almost linearly. This $W^{*}$ is found to be proportional to $L^{-1/\nu}$ with \textcolor{black}{$\nu=0.63(3)$}, which is obtained via fitting as shown using the solid line in (b), implying the onset of localization in the thermodynamic limit even at infinitesimal disorder. (c) illustrates localization length at $W^*$, $\xi_g^*$, as a function of $L$. The circles stand for the data points, and the solid line is the best fit. It turns out that $\xi_g^* \propto L$ which suggests $\mu=-1$ in Eq.~\ref{Collapse}.  (d) depicts the collapse plot, where different plots corresponding to different system sizes in (a)  merge for \textcolor{black}{$\nu=0.63(1)$}. In both (a) and (d), horizontal and vertical axes are in log$_{10}$ scale.}
    \label{Localization Length}
\end{figure}

We compute the localization length $\xi_g$ for the ground state as a function of the disorder strength $W$.  Fig.~\ref{Localization Length}(a) illustrates the cases with various system sizes $L$. There is a prominent trend that one may follow: For a given system size, the ground state localization length, $\xi_g$, remains almost flat upto a certain disorder strength, say $W^*$, beyond which it starts decaying. We would like to mention that in this manuscript we call $W^*$ the transition point for the finite systems, and use a different symbol $W_c$ in the thermodynamic limit to distinguish it from the finite-size systems, i.e., $W^*\equiv W_c$ as $L\to \infty$. As shown in Fig.~\ref{Localization Length}(a), it can be observed that the flat region starts shrinking with the increase of the system size. One may anticipate the vanishing of the flat region with $W^*$ tending to zero as the system approaches the thermodynamic limit. This would, in turn, imply the onset of localization for infinitesimally small disorder strength at infinite system size, i.e., $W^*\equiv W_c \to 0$ as $L \to \infty$, and as discussed before, $\xi_g$ is supposed to obey the suggested algebraic scaling law controlled via the scaling exponent, $\nu$. To extract $\nu$ from the finite-size scaling analysis, one, however, needs to perform a careful analysis via the rescaled variables. This can be done by proceeding in accordance with the conventional second-order finite-size scaling analysis via the well-known methodology of data collapse. 

We consider the following scaling ansatz for the localization length for the ground state: 
\begin{equation}
    L^{\mu}\xi_g(W)=f\left[L^{(1/\nu)}\left|W-W_c\right| \right],
    \label{Collapse}
\end{equation}
where $f[\cdot]$ is an arbitrary function. This scaling ansatz is verified via data collapse. The idea of the data collapse is the following: We obtain data corresponding to $L^{\mu}\xi_g(W)$ versus $L^{(1/\nu)}\left|W-W_c\right|$ for different system sizes $L$. 
The parameters $\mu$, $\nu$, and $W_c$ are then needed to be tuned so that different data corresponding to different sizes collapse on top of each other.  Moreover, the unbiased best-fit values of the scaling exponents and the critical point are obtained from the data collapse via the cost function approach \cite{Suntajs20}.

{\color{black}In order to get a rough idea for $W_c$, we look at how the $W^*$ value for each system size scales as a function of $L$ and by doing a fitting in Fig.~\ref{Localization Length}(b) we get $W^*\propto L^{-1/\nu}$ with $\nu=0.63(3)$ and this indicates that in the thermodynamic limit $W^*\equiv W_c\to 0$. Also, to get an idea of how the ground state localization length $\xi_g^*$ near $W^*$  depends on $L$, we do a fitting of $\xi_g^*$ as a function of $L$ in Fig.~\ref{Localization Length}(c), which shows: $\xi_g^*\propto L$ near $W^*$, suggesting that $\mu=-1$. Making use of this knowledge, we acquire the fine-tuned values of the parameters $W_c$ and $\nu$ through cost function minimization (see appendix) by choosing $Q=\xi_g /L$, which suggests $\nu=0.63(1)$, and $W_c\to0$. We present the corresponding collapse plot in Fig.~\ref{Localization Length}(d). We have noticed that if we use system sizes $L=100, 200, 500, 750, \text{and}\, 1000$ instead, we get $\nu=0.65(1)$. Looking at this trend, we infer that for even larger system sizes, this $\nu$ value will take up the known $2/3$ value for the 1D Anderson localization and thus this model belongs to the same universality class as the Anderson model.}

Moreover, the nature of the crossover can be unveiled via finite-size scaling analysis of other suitable quantities, such as inverse participation ratio (IPR), or the average energy splitting between two consecutive levels, $\Delta E$. The inverse participation ratio (IPR) of any eigenstate is defined as,
\begin{equation}
    I=\sum_{j=1}^{L}p_j^2.
\end{equation}
\begin{figure}[t]
    \centering
    \includegraphics[width=\linewidth]{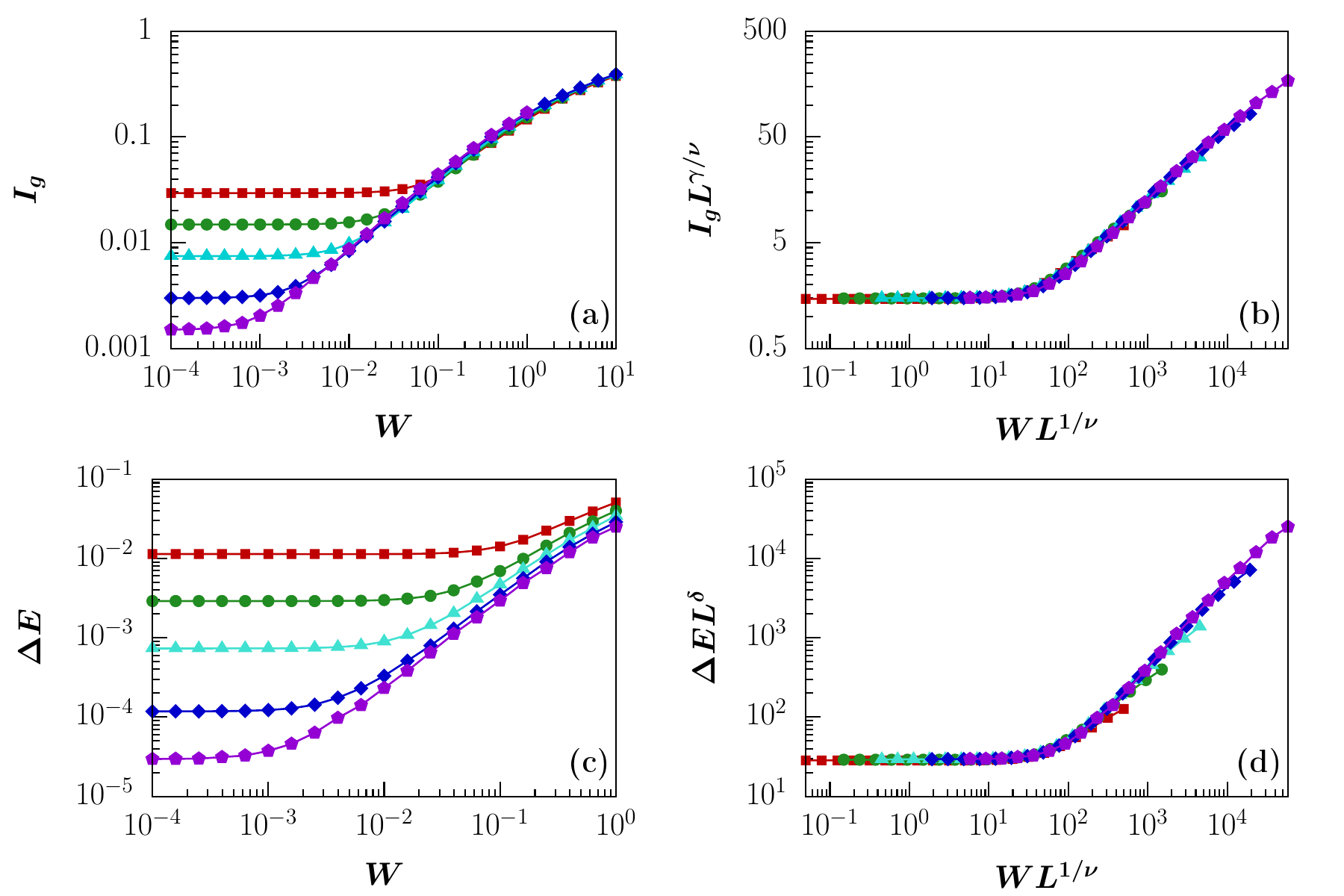}
    \caption{{\bf Inverse Participation Ratio and Energy Splitting:} The IPR for the ground state, $I_g$, is plotted as a function of $W$ in (a) for different $L$. Here also, as in the case of localization length, there is a flat region in $I_g$ upto the same $W^*$ for each system size that scales as $W^*\propto L^{-1/\nu}$. The corresponding collapse plot is shown in (b), where the perfect data collapse is found at $\gamma=\nu=0.63(1)$. This suggests that  $I_g\propto L^{-1}$ near $W_c$. The average level spacing $\Delta E$ between the ground state and the $1$st excited state remains small for small values of $W$ and after a certain $W^*$ for each system it continues to increase as shown in (c). From the scaling analysis in (d) it is found that $\Delta E$ scales as $L^{-\delta}$ with $\delta=2$. This implies that although the system is gapless in the thermodynamic limit, the introduction of infinitesimal disorder opens up a gap. The same point-type scheme is used here as in Fig.~\ref{Localization Length} to indicate different system sizes. For both the plots horizontal and vertical axes are in log$_{10}$ scale.}
    \label{IPR_delE}
\end{figure}
Here we study the variation in ground state IPR, $I_g$, as a function of $W$ for different $L$'s. 
Fig. \ref{IPR_delE}(a) represents the variations of IPR with the disorder strength for different system sizes. As one may expect, the IPR is characterized by a nearly flat region upto $W^*$ for a given system size, beyond which it grows roughly linearly with respect to $W$. The extension of the flat region upto $W^*$ diminishes with increasing system size, the same as the localization length. Following the analysis presented in the context of the localization length, the vanishing of the $W^*$ in the thermodynamic limit can be safely predicted, i.e., $W^* \equiv W_c\to 0$ for $L\to \infty$. The IPRs corresponding to different system sizes merge at large $W$, highlighting the localized nature of the wavefunction. Considering that the behaviour of the IPR is controlled via another scaling exponent, $\gamma$, such that IPR scales as $I_g \propto W^{\gamma}$ near $W_c$, we perform a scaling analysis analogous to the case of the localization length. We rescale $W$ as $WL^{1/\nu}$ and $I_g$ as $I_gL^{\gamma/\nu}$, implying the following scaling ansatz: 
\begin{equation}
    L^{(\gamma/\nu)} I_g(W)= g\left[L^{(1/\nu)}\left|W-W_c\right| \right],
\end{equation}
where $g$[$\cdot$] is again an arbitrary function. Confirmation of this scaling ansatz is achieved through the collapse plots in Fig.~\ref{IPR_delE}(b) corresponding to different $L$'s.  In order to extract $\gamma$, we again adopt the cost function minimization approach. $\gamma$ turns out to be quantitatively equal to $\nu$, i.e., $\gamma=\nu=0.63(1)$. One may infer the scaling of IPR with the system size near the crossover as $I_g\propto L^{-\gamma/\nu}$. It's worth mentioning that $\gamma$ is equal to $\nu$ in the case of the pure Anderson model as well, both of which are equal to $2/3$, but for the pure AA model $\nu=1.0$ and $\gamma\approx0.33$ \cite{Bu22}.

Finally, we study the energy splitting in order to characterize the crossover. For this, we consider the energy gap, $\Delta E$, between the ground state and 1st excited state energy. In order to perform the finite-size scaling for $\Delta E$: 
\begin{equation}
    L^{\delta} \Delta E(W)= h\left[L^{(1/\nu)}\left|W-W_c\right| \right],
\end{equation}
where we aim to justify the ansatz via data collapse, and then to find the scaling exponent, $\delta$, from thereof. Fig. \ref{IPR_delE}(c) shows the energy gap $\Delta E$ as a function of $W$. Near $W_c$, in the thermodynamic limit, the energy gap scales as $\Delta E \propto L^{-\delta}$. The scaling exponent $\delta$ is obtained from data collapse, where the rescaling for $W$ is the same as before, and the rescaling for $\Delta E$ is done as $\Delta E L^\delta$. The collapse plot corresponding to $\Delta E$ is presented in Fig.~\ref{IPR_delE}(d).  Interestingly, this scaling exponent comes out to be the same as what one finds in the case of Anderson localization, i.e., $\delta\approx2$. In the thermodynamic limit, the scaling of $\Delta E$ with respect to $W$ is given by $\Delta E \propto W^{\nu \delta}$.

\section{Discussion}
{\color{black}In this work, we have used a tight-binding model with random nearest-neighbor tunnelling without any on-site potential. The hopping strength assumes real values, but the left-to-right ones are different from the right-to-left ones and it varies randomly from site to site.  The random hopping strength has been chosen in such a fashion that the sign of hopping strength between the adjacent lattice, i.e., from left-to-right and from right-to-left have the same sign. Under such a circumstance the resulting Hamiltonian has an entirely real spectrum, although the Hamiltonian itself is non-Hermitian in nature. 

The disorder in hopping results in localization in the system. The well-known Hatano-Nelson model is also a model with off-diagonal disorder but the disorder is not at all random, meaning that there is some preferred direction of movement of the particle and for that reason, the wavefunctions get localized at that edge of the lattice. In contrast to that, in our model, there is no such preferred direction and likewise, the wavefunctions get localized at different positions of the lattice.

In this work first, we have conducted a study on the spectral statistics, which has come up with the result that the spectral statistics remains Wigner-Dyson type at small disorder and becomes Poissonian at large disorder strength. 
In between, there is a continuous change in spectral statistics from the Wigner-Dyson type to the Poisson distribution. For the band edge of the energy spectrum, the delocalization-localization crossover starts at smaller disorders compared to the mid-band states. However, at the center of the spectrum, the wavefunction is always delocalized which is related to the singularity in the DOS at the mid-band. Band-edge levels exhibit Poisson statistics already at parameter values where mid-band levels still display Wigner-Dyson statistics. This separation
of crossover scales, quantified in Fig. 1, provides a spectral counterpart to the mid-band delocalized state and DOS singularity.

While the existence of an $E=0$ delocalized state is known in Hermitian chiral Anderson models, our work demonstrates
how this critical state is modified or persists under non-Hermitian off-diagonal disorder, a direction that has received no systematic scaling analysis. In contrast to the previous works, which treat Hermitian hopping disorder, our model realizes the mid-band delocalized state and DOS singularity in a non-Hermitian but pseudo-Hermitian setting. By enforcing the sign constraint on the asymmetric hoppings, we explicitly connect the random-hopping problem to a class of non-Hermitian tridiagonal matrices whose spectra
are rigorously real. This connection between off-diagonal localization and pseudo-Hermitian random matrices is, to our knowledge, absent from the earlier literature. We analyze the right eigenvectors and find that they exhibit delocalized behaviour; from
test cases we understood left eigenvectors show similar features when configurational averaging is performed over large sample size. One might argue that this delocalization simply
reflects properties of the Hermitian Hamiltonian obtained through the mapping. However, it is important to note that while a similarity transformation guarantees
only that the spectrum is preserved. It does not preserve the structure of the
eigenvectors, and hence, does not guarantee preservation of localization properties and
universality class at criticality of the system apriori, without performing a detailed study. Therefore, the emergence of delocalized eigenvectors directly in the non-Hermitian setting is a nontrivial result.

{\color{black}In order to unearth the nature of the system near $W_c$, we have done a careful finite-size scaling analysis via  localization length, IPR, and the average spacing between the ground state and the first excited state energy. It has been found that for a particular system size $L$, after a certain disorder $W^*$, the system continuously transits from the delocalized to the localized state and this $W^*$ is proportional to $L^{1/\nu}$, where $\nu=0.65(1)$, and thus indicates the universality class of the pure Anderson localization, for which $\nu=2/3$. Similar to the Anderson localization, in the thermodynamic limit i.e., as $L \to \infty$, $W^* \equiv W_c \to 0$. As in the case of traditional Anderson localization here also we have found that the localization length for the ground state, $\xi_g^* \propto L$ at $W^*$, while it is $I_g^* \propto L^{-1}$ for IPR of the ground state and $\Delta E^*\propto L^{-2}$ for average level spacing $\Delta E$ between ground and first excited state.  It is nevertheless non-trivial to expect a priori, before performing detailed analysis, that the same exponent controls localization length, IPR, and gap scaling
also in this pseudo-Hermitian realization of off-diagonal disorder. We stress upon the fact that our work provides first numerical evidence for a non-Hermitian extension of the chiral universality class.To our knowledge, such a systematic
scaling characterization has not been reported previously for one-dimensional off-diagonal disorder.}

Finally, we would like to remark on the close connection to experimentally accessible circuit platforms that makes our results particularly timely and relevant.
The Hamiltonian studied here can be naturally realized in electrical circuit networks,
where the circuit Laplacian acts as an effective tight-binding Hamiltonian with tunable non-Hermitian couplings. Asymmetric and randomly distributed nearest-neighbor hoppings can
be engineered using active circuit elements, while maintaining a uniform sign of the hopping
products ensures a purely real spectrum, as required in our model. The localization proper1ties and spectral features discussed in this work can be directly probed through impedance
or voltage-response measurements in such circuits, which have recently been used to realize
non-Hermitian lattice models \cite{Zhang}.}

\section*{Ackownledgement}
D.R. acknowledges support from the Science and Engineering Research Board (SERB), Department of Science and Technology (DST), under the sanction No. SRG/2021/002316-G.

\section*{Appendix}

{\color{black}Considering $N_p$ data values for a given quantity under study, say $Q$, for different $W$ and $L$, the cost function, $C_Q$, is defined as,
\begin{equation}
    C_Q = \frac{\sum_i^{N_p-1} |Q_{i+1}-Q_i|}{\max(Q_i)-\min(Q_i)} - 1.
    \label{Cost_Function}
\end{equation}
One needs to sort all $N_p$ values of $|Q_i|$ according to non-decreasing values of $L~\textrm{sgn} [W-W_c]|W-W_c|^{\nu}$. The perfect data collapse would correspond to a situation where $\sum_{i}^{N_p-1} |Q_{i+1}-Q_i| = \max(Q_i)-\min(Q_i)$, implying $C_Q=0$. Otherwise, $C_Q$ assumes a positive value.

\begin{figure}[t]
    \centering
    \includegraphics[width=\linewidth]{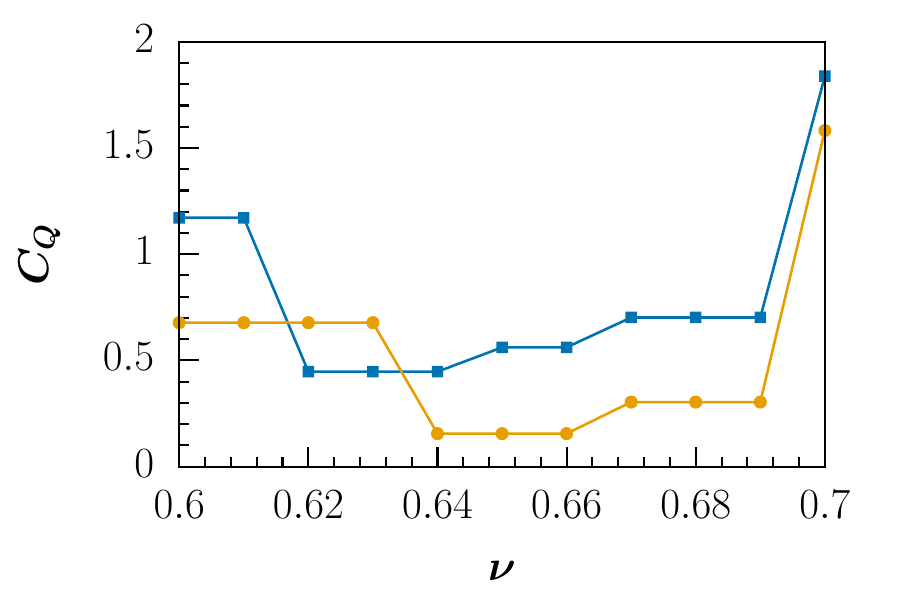}
    \caption{{\color{black}The blue squares and orange circles correspond to the cost function values for the collapse of system sizes $L=50,100,200,500,1000$ and $L=100,200,500,750,1000$ respectively. The solid lines are guides to the eye.}}
    \label{fig:Cost_Fn}
\end{figure}

In reality, one never gets a perfect collapse. So the strategy is to change the $\nu$ value gradually and see where the global minimum is. In our case, we find a plateau of global minima from where we choose the mid-point as the $\nu$ value with an errorbar as half of the length of the plateau. For the blue squares in Fig.~\ref{fig:Cost_Fn}, the minima are found between $0.62$ and $0.64$, so we take $0.63(1)$. On the other hand, for the orange circles, the minima are between $0.64$ and $0.66$, so we take $0.65(1)$ as the $\nu$ value. This suggests that larger system sizes are better for the extraction of the scaling exponent via the data collapse technique.}

\end{document}